\documentclass[twocolumn]{aastex631}

\graphicspath{{./}{figures/}}
\usepackage{amsmath}

\usepackage{CJK}

\newcommand\sref[1]{\hyperref[#1]{\S~\ref*{#1}}}
\newcommand\fref[1]{\hyperref[#1]{Fig.~\ref*{#1}}}
\newcommand\Eqref[1]{Eq.~(\hyperref[#1]{\ref*{#1}})}
\newcommand\eeqref[1]{Eq.~\hyperref[#1]{\ref*{#1}}}
\newcommand\tref[1]{\hyperref[#1]{Table~\ref*{#1}}}
\newcommand\aref[1]{\hyperref[#1]{Appendix~\ref*{#1}}}

\newcommand{\Msun}{{\rm M_\odot}}

\shorttitle{Bridging Scales: Galactic nucleus to cosmic environment}
\shortauthors{Su et al.}

\begin{document}
\begin{CJK}{UTF8}{mj}
\title{Bridging Scales: Coupling the galactic nucleus to the larger cosmic environment}

\author[0000-0003-1598-0083]{Kung-Yi Su}
\affiliation{Black Hole Initiative at Harvard University, 20 Garden Street, Cambridge, MA 02138, USA}
\affiliation{Center for Astrophysics $\vert$ Harvard \& Smithsonian, 60 Garden Street, Cambridge, MA 02138, USA}

\author[0000-0002-5554-8896]{Priyamvada Natarajan}
\affiliation{Black Hole Initiative at Harvard University, 20 Garden Street, Cambridge, MA 02138, USA}
\affiliation{Department of Astronomy, Yale University, Kline Tower, 266 Whitney Avenue, New Haven, CT 06511, USA}
\affiliation{Department of Physics, Yale University, P.O. Box 208121, New Haven, CT 06520, USA}

\author[0000-0002-2858-9481]{Hyerin Cho (조혜린)}
\affiliation{Center for Astrophysics $\vert$ Harvard \& Smithsonian, 60 Garden Street, Cambridge, MA 02138, USA}
\affiliation{Black Hole Initiative at Harvard University, 20 Garden Street, Cambridge, MA 02138, USA}

\author[0000-0002-1919-2730]{Ramesh Narayan}
\affiliation{Center for Astrophysics $\vert$ Harvard \& Smithsonian, 60 Garden Street, Cambridge, MA 02138, USA}
\affiliation{Black Hole Initiative at Harvard University, 20 Garden Street, Cambridge, MA 02138, USA}

\author[0000-0003-3729-1684]{Philip F. Hopkins}
\affiliation{TAPIR, Mailcode 350-17, California Institute of Technology, Pasadena, CA 91125, USA}

\author[0000-0001-5769-4945]{Daniel Angl{\'e}s-Alc{\'a}zar}
\affiliation{Department of Physics, University of Connecticut, 196 Auditorium Road, U-3046, Storrs, CT 06269, USA}

\author[0000-0002-0393-7734]{Ben S. Prather} 
\affiliation{CCS-2, Los Alamos National Laboratory, PO Box 1663, Los Alamos, NM 87545, USA}

\begin{abstract}\label{abstract}
Coupling black hole (BH) feeding and feedback involves interactions across vast spatial and temporal scales that is computationally challenging. Tracking gas inflows and outflows from kilo-parsec scales to the event horizon for non-spinning BHs in the presence of strong magnetic fields, \cite{Cho+2023, Cho+2024} report strong suppression of accretion on horizon scales and low (2\%) feedback efficiency. In this letter, we explore the impact of these findings for the supermassive BHs M87* and Sgr A*, using high-resolution, non-cosmological, magnetohydrodynamic (MHD) simulations with the Feedback In Realistic Environments (FIRE-2) model. With no feedback, we find rapid BH growth due to ``cooling flows,'' and for 2\% efficiency feedback, while accretion is suppressed, the rates still remain higher than constraints from Event Horizon Telescope (EHT) data \citep{EHTm87_8_2021, EHTMW_5_2022} for M87* and Sgr A*. To match EHT observations of M87*, a feedback efficiency greater than 15\% is required, suggesting the need to include enhanced feedback from BH spin. Similarly, a feedback efficiency of $>15\%$ is needed for Sgr A* to match the estimated observed star formation rate of $\lesssim 2 {\rm M_\odot}$ yr$^{-1}$. However, even with 100\% feedback efficiency, the accretion rate onto Sgr A* matches with EHT data only on rare occasions in the simulations, suggesting that Sgr A* is likely in a temporary quiescent phase currently. Bridging accretion and feedback across scales, we conclude that higher feedback efficiency, possibly due to non-zero BH spin, is necessary to suppress ``cooling flows'' and match observed accretion and star formation rates in M87* and Sgr A*.
\end{abstract}

\section{Introduction}
\label{S:intro}

The nature of physical processes that link the feeding of supermassive BHs to their feedback effects on cosmological scales is still an open question and under investigation. Feedback from accreting BHs, AGN, is believed to quench star formation in massive galaxies, keeping them in a ``red and dead" state over extended cosmic epochs in concordance with observations. However, understanding how this occurs remains a challenge for studies of galaxy evolution in high-mass halos with $M \gtrsim 10^{12}M_{\odot}$ \citep[e.g.,][]{Bell+2003,Kauffmann+2003,Madgwick+2003,Baldry2004,keres+2005,Blanton+2005,Dekel+2006,keres+2009,Pozzetti+2010,Wetzel+2012,Feldmann+2015,Voit+2015}. Multi-wavelength observations, especially X-rays, reveal the presence of gas in the cores of massive elliptical galaxies and clusters, that could potentially fuel star formation \citep[e.g.,][]{Fabian+1994,Peterson+2006,McDonald+2011,Werner+2013,Stern+2019,Mercedes+2023}. However, this expected star formation is not observed \citep{Tamura+2001,ODea+2008,Rafferty+2008}, leading to the well recognized ``cooling flow problem.'' Star formation appears to be curtailed in the galactic nuclei of massive galaxies. Non-AGN quenching mechanisms, such as stellar feedback or cosmic rays, appear to be ineffective \citep{Su+2019} to account for quenching, implicating AGN feedback.

However, understanding of how AGN feedback operates is limited due to the lack of first-principles physics, sparse observational constraints, and computational limitations. Cosmological simulations on galaxy scales, therefore, rely on sub-grid AGN feedback and BH accretion models, that utilize scaling laws to implement astrophysical processes that lie beyond current numerical resolution limits \citep[e.g.,][]{Li2014,Fiacconi2018,AnglesAlcazar+Zoom_2021,Talbot2021,Weinberger2023}. Therefore, sub-grid recipes that approximate physics are commonly used in all major simulation suites \citep[e.g.,][]{Sijacki2015,Rosas-Guevara2016,AngleAlcazar+2017,Weinberger2018,Ricarte2019,Ni2022,Wellons2023}. Meanwhile, general relativistic magnetohydrodynamic (GRMHD) simulations, which resolve horizon-scale BH physics, are able to self-consistently launch jets, highlighting a potential feedback mechanism \citep[e.g.,][]{Gammie2003, Tchekhovskoy2011, Porth:2019, Chatterjee:2023}. Yet, these simulations are still idealized, as they do not take the larger cosmological context into account due to the extensive computational resources required. As a result, gaps remain in our understanding of AGN feedback.

\citet{Cho+2023, Cho+2024} introduced a new multi-zone method to successfully couple the disparate physical scales relevant to BH accretion and AGN feedback. This multi-zone approach partitions spatial scales from the BH event horizon to that of the galaxy into several annuli and at any given time, only one of these annuli is actively evolved while the others remain frozen, which alleviates the prohibitive time-step issue making the calculation computationally tractable. By iteratively cycling through the active annuli up and down the radial scale several hundred times, for the first time, a converged, dynamical steady-state solution for both inflow and outflow processes that connect physical scales spanning over seven orders of magnitude was achieved. 

The first application of this multi-zone method to magnetized Bondi accretion \citep{Cho+2023} revealed that continuous feedback occurs over a vast dynamic range in spatial scales even for non-spinning BHs. The feedback geometry was quasi-spherical with a magnitude of about $\approx 0.02\, \dot{M}_{\rm BH}c^2$, likely powered by convection triggered by magnetic reconnection in the vicinity of the BH. Additional findings include the suppression of the accretion rate at the horizon by two orders of magnitude compared to the Bondi estimate $\dot{M}_{\rm BH}\approx 0.01 \dot{M}_B$ and a density scaling of $\rho\propto r^{-1}$, both consistent with EHT observations of M87* and Sgr A*. The plasma-$\beta$ was found to be saturated at order unity over many decades in radius, indicating the presence of a magnetically arrested disk. \citet{Cho+2024} followed up with a more realistic treatment of the galactic environment by adopting initial conditions and the gravitational potential from an isolated galaxy simulation. With these boundary conditions, key results of the steady state remained consistent with the idealized magnetized Bondi accretion case with mild feedback extending over seven orders of magnitude out from the event horizon. 

In this letter, we present the results of the first attempt to directly bridge galaxy simulations incorporating the findings from the extended GRMHD simulations. We specifically explore and quantify the impact of suppressed Bondi accretion and AGN feedback seen in the multi-zone GRMHD simulations on significantly larger scales. Additionally, in this work, we also examine results from runs in which the AGN feedback efficiency is varied to higher values, as expected for spinning BHs. BH spin has not yet been incorporated into the multi-zone GRMHD simulations at present.

\section{Incorporating galaxy scales: Methodology} \label{S:methods}

We utilize isolated galaxy simulations to simulate M87* and Sgr A* like systems using the {\small GIZMO}\footnote{A public version of  GIZMO is available at \href{http://www.tapir.caltech.edu/~phopkins/Site/GIZMO.html}{\textit{http://www.tapir.caltech.edu/$\sim$phopkins/Site/GIZMO.html}}} code \citep{2015MNRAS.450...53H}. For these runs, we use {\small GIZMO} in its mesh-less finite mass (MFM) mode which uses a Lagrangian mesh-free Godunov method for hydrodynamics, capturing both the advantages of grid-based and smoothed-particle hydrodynamics (SPH) methods. Numerical details and extensive tests of this code are described in a series of previously published methods papers for, e.g.,\ the hydrodynamics and self-gravity \citep{2015MNRAS.450...53H} and magnetohydrodynamics \citep[MHD;][]{2016MNRAS.455...51H,2015arXiv150907877H}.

\begin{table*}
\begin{center}
 \caption{Properties of Initial Conditions in our Simulations}
 \label{tab:ic}
 \resizebox{18.cm}{!}
 {
 \movetableright=-2.2in
 \begin{tabular}{l|C| c c |c|c c c| c |c c| c c| c c| c c c }
 \hline
\hline
&&\multicolumn{2}{c|}{\underline{Resolution}}&\underline{BH}&\multicolumn{3}{c|}{\underline{DM halo}}&&\multicolumn{2}{c|}{\underline{Stellar Bulge}}&\multicolumn{2}{c|}{\underline{Stellar Disc}}&\multicolumn{2}{c|}{\underline{Gas Disc}}&\multicolumn{3}{c}{\underline{Gas Halo}}  \\
$\,\,\,\,$Model &R_{200} &$\epsilon_g$ &$m_g$        &$M_{\rm BH}$&$M_{\rm halo}$   &$r_{dh}$            &$\rho_{0}$    &$M_{\rm bar}$    &$M_b$ 
                 &$a$          &$M_d$        & $r_d$             &$M_{gd}$       &$r_{gd}$         &$M_{gh}$         &$r_{gh}$ &$\beta$    \\
                &(kpc) &(pc)        &(M$_\odot$) &(M$_\odot$)  &(M$_\odot$)      & (kpc)           &(g cm$^{-3}$)           &(M$_\odot$)      &(M$_\odot$) 
                  &(kpc)        &(M$_\odot$)  &(kpc)            &(M$_\odot$)    &(kpc)            &(M$_\odot$)      &(kpc)&\\
\hline
$\,\,\,\,$M87*        &1306&1       &2e4   &6.5e9        &2.0e14           &448             &807              &5.8e13           &6.9e11   
                     &3.97       &-          &-                &-            &-                &5.1e13           &0.93   &0.33     \\  
$\,\,\,\,$Sgr A*        &248&1       & 8e3 &4.3e6          &1.5e12           &20             &174              &2.3e11           &1.5e10   
                     &1.0       &5.0e10          &3.0                &5.0e9            &6.0                &1.6e11           &20  &0.5      \\  
$\,\,\,\,$Sgr A* -light CGM  &241&1       & 8e3 &4.3e6          &1.5e12           &20             &174              &1.2e11           &1.5e10   
                     &1.0       &5.0e10          &3.0                &5.0e9            &6.0                &4.6e10           &20  &0.5      \\

\hline 
\end{tabular}

}
\resizebox{18.cm}{!}
 {
  \movetableright=-1.56in
\begin{tabular}{lc|cc|cccccc|cccc}

\hline
&&\multicolumn{2}{c|}{Numerical details} & \multicolumn{6}{c|}{Feedback parameter} & \multicolumn{4}{c}{Results } \\
\hline
\multicolumn{2}{c|}{Model} & $\Delta T$ & $m_{\rm wind}$  &  $T_{\rm wind}$&  $B_{\rm wind}$ &  $V_{\rm wind}$ & $f_{\rm mass, acc}$ & $f_{\rm mass, wind}$ & $\eta_{\rm Kin}$  & $M^{\rm 1.5 Gyr}_{\rm BH}$ &$M^{\rm 1.5 Gyr}_{\rm *}$ & $\langle\dot{M}_{\rm BH}\rangle$ & $\langle\dot{M}_{\rm *}\rangle$ \\
&&(Gyr) & (${\rm M_\odot}$) & (K) & (G)& (km s$^{-1}$) &  &  &  & (${\rm M_\odot}$) & (${\rm M_\odot}$)& (${\rm M_\odot}$ yr$^{-1}$) & (${\rm M_\odot}$ yr$^{-1}$) \\
\hline
\multicolumn{2}{l|}{\underline{\textcolor{blue}{M87*}}}&&&&&&&&\\
A&M87*-Bondi           &1.5& -   & - &- &  -   &1   &   0&0&2.8e10&6.9e11&17 &0\\
B&M87*-subBondi        &1.3& -   & - &- &  -   &0.01&   0&0&2.1e10&6.9e11&14 &1.3\\
C1&M87*-subBondi-efflow &1.5&1.8e4&1e7&1e-4&6e3   &0.01&0.99&0.02&6.6e9&6.9e11&6.0e-2&0.47\\
C2&M87*-subBondi-effmed &1.5&1.8e4&1e7&1e-4&1.65e4&0.01&0.99&0.15&6.5e9&6.9e11&4.9e-3&0.12\\
C3&M87*-subBondi-effhigh&1.5&1.8e4&1e7&1e-5&4.2e4 &0.01&0.99&1   &6.5e9&6.9e11&3.0e-4&7.9e-2\\

\hline
\multicolumn{2}{l|}{\underline{\textcolor{blue}{Sgr A*}}}&&&&&&&&&\\
A&SgrA*-Bondi           &2.5& -   & - &- &  -   &1   &   0&  0      &7.4e9&6.5e10&5.0   &9.3e-2\\
B&SgrA*-subBondi        &3  & -   & - &- & -     & 0.01&  0 &0     &3.0e9&6.6e10&2.3   &0.46\\
C1&SgrA*-subBondi-efflow &3&8e3&1e7&1e-4&6e3   &0.01&0.99&0.02      &4.6e6&7.0e10&3.2e-4&4.3\\
C2&SgrA*-subBondi-effmed &3&8e3&1e7&1e-4&1.65e4&0.01&0.99&0.15      &4.4e6&6.9e10&6.7e-5&3.9\\
C3&SgrA*-subBondi-effhigh&3&8e3&1e7&1e-4&4.2e4 &0.01&0.99&1         &4.3e6&6.6e10&1.8e-5&1.0\\
\hline
\multicolumn{2}{l|}{\underline{\textcolor{blue}{Sgr A* -light CGM}}}&&&&&&&&&\\
A&SgrA*-lightCGM-Bondi           &3& -   & - &- &  -   &1   &   0&  0  &2.9e9&6.5e10&0.97  &7.8e-2\\
C1&SgrA*-lightCGM-subBondi-efflow &3&8e3&1e7&1e-4&6e3   &0.01&0.99&0.02&4.6e6&6.7e10&4.2e-4&1.0\\
C3&SgrA*-lightCGM-subBondi-effhigh&3&8e3&1e7&1e-4&4.2e4 &0.01&0.99&1   &4.3e6&6.6e10&1.1e-5&0.33\\

\hline 
\hline
\end{tabular}
}
\end{center}
\begin{flushleft}
{\bf Upper} -- Parameters of the galaxy models studied here : 
(1) Model name. 
(2) $R_{200}$.
(3) $\epsilon_g$: Minimum gravitational force softening for gas (the softening for gas in all simulations is adaptive, and matched to the hydrodynamic resolution; here, we quote the minimum Plummer equivalent softening).
(4) $m_g$: Gas mass (resolution element). There is a resolution gradient for M87* analog runs, 
so its $m_g$ is the mass of the highest resolution elements.
(5)$M_{\rm BH}$: BH mass.
(6) $M_{\rm halo}$: Dark matter halo mass within $R_{\rm vir}$. 
(7) $r_{dh}$: NFW halo scale radius, which yields a central concentration parameter
$c \sim$ 3 [MW halo] \& 12 [for the M87 halo].
(8) $\rho_{\rm 0}$: The density normalization for the NFW profile.
(9) $M_{\rm bar}$: Total baryonic mass within $R_{\rm vir}$. 
(10) $M_b$: Bulge mass.
(11) $a$: Bulge Hernquist-profile scale-length.
(12) $M_d$ : Stellar disc mass.
(13) $r_d$ : Stellar disc exponential scale-length.
(14) $M_{gd}$: Gas disc mass. 
(15) $r_{gd}$: Gas disc exponential scale-length.
(16) $M_{gh}$: Hydrostatic gas halo mass within $R_{\rm vir}$. 
(17) $r_{gh}$: Hydrostatic gas halo $\beta$ profile scale-length.
(18) $\beta$: Hydrostatic gas halo $\beta$ profile $\beta$.\\
{\bf Lower} -- List of the runs: 
(1)Model name. 
(2) $\Delta T$: Duration of the runs. 
(3) $m_{\rm wind}$: Mass resolution of the spawned AGN wind particles. 
(4,5,6) $T_{\rm wind}$, $B_{\rm wind}$, and  $V_{\rm wind}$: Initial temperature, magnetic field strength, and velocity of the wind particles. 
(7,8) $f_{\rm mass, acc}$ $f_{\rm mass, wind}$: The accretion rate and the wind mass flux with respect to Bondi. 
(9) $\eta_{\rm Kin}$: The  kinetic AGN feedback efficiencies. 
(10) $M^{\rm 1.5 Gyr}_{BH}$: BH mass and stellar mass after 1.5 Gyr. 
(11) $M^{\rm 1.5 Gyr}_{*}$:  The averaged star formation rates up to 1.5 Gyr for M87* analog runs and up to 2.5 Gyr for Sgr A* analog runs.
(12,13) $\langle\dot{M}_{\rm BH}\rangle$  $\langle\dot{M}_{\rm *}\rangle$: The averaged BH accretion rate from 0.5-1.5 Gyr for M87* analog runs and from 0.5-2.5 Gyr for Sgr A* analog runs.

\end{flushleft}
\end{table*}

All our simulations adopt the FIRE-2 implementation of the  treatment of the ISM, star formation, and stellar feedback, as summarized in \aref{a:sfb}. The details of this well tested and extensively used code are provided in \citet{Hopkins+FIRE2018} along with numerical tests. Here we provide a brief summary of the essentials. In FIRE-2 gas cooling is followed from 10-$10^{10}$ K and stellar feedback includes supernovae, stellar winds, and radiative feedback therefrom. Star formation is treated via a sink particle method, allowing only dense, molecular, self-shielding, locally self-gravitating gas to form stars. In addition, the simulation implements BH accretion following the standard Bondi prescription, accounting for the sound speed and the relative velocity between the BH and the surrounding gas.
\begin{align}\label{eq:bondi}
\dot{M}_{\rm Bondi}=\frac{4\pi\rho G^2 M_{\rm BH}^2}{(c_s^2+\Delta v^2)^{3/2}}\notag\\
\dot{M}_{\rm BH}=f_{\rm mass, acc} \dot{M}_{\rm Bondi},
\end{align}
where $c_s$ is the averaged sound speed and $\Delta v$ is the averaged relative velocity around the BH. The accretion rate on the BH is proportional to the Bondi rate via the pre-factor $f_{\rm mass, acc}$ shown above. AGN feedback is modeled as an isotropic wind and 
the energy of the wind is predominantly kinetic, and the corresponding outflow mass and energy flux are given by:
\begin{align}
\dot{M}_{\rm wind}=f_{\rm mass, wind} \dot{M}_{\rm Bondi}.\notag\\
\dot{E}_{\rm wind} = \frac{1}{2} \dot{M}_{\rm wind} V_{\rm wind} ^2.
\end{align}
Therefore the feedback efficiency, $\eta \equiv \dot{E}_{\rm wind}/\dot{M}_{\rm BH} c^2$, can be written as:
\begin{align}
\eta\sim\eta_{\rm kin}\equiv \frac{f_{\rm mass, wind} V_{\rm wind} ^2}{2 f_{\rm mass, acc}c^2},
\end{align}
with $f_{\rm mass, acc}+f_{\rm mass, wind}=1$.
The numerical implementations are summarized in \aref{a:agn}. The neglected geometric effects and other limitations of the model are briefly discussed in \aref{a:numerical} and are deferred for future study.

We conduct the following five sets of runs: (i) Run A - growth with full Bondi accretion without AGN feedback; (ii) Run B - suppressed BH accretion with respect to Bondi ($f_{\rm mass, acc}=0.01$) without AGN feedback; and (iii) Runs C1, C2 and C3 that all deploy suppressed Bondi accretion with three different AGN feedback efficiencies from low to high: $\eta =$ 0.02, 0.15, 1. Detailed parameters characterizing each of these runs are outlined in \tref{tab:ic}.
 
Initial conditions for the galaxy-scale simulations are adopted to resemble an M87*-like host galaxy and associated halo \citep{2008MNRAS.388.1062C, 2012MNRAS.421..635F, 2016MNRAS.457..421O}, and the Sgr A*- host Milky Way halo \citep{Miller+MW2013, Miller+MW2015, Gupta+2017}. Additionally, we explore two sets of initial conditions for the Sgr A* analog, that correspond to different Circum-Galactic Medium (referred to as CGM) masses (i) utilizing the CGM profile from \cite{Miller+MW2015}, referred to as Sgr A*-light CGM, consistent with missing baryons, and (ii) using a similar profile with a universal baryonic fraction (0.16) within twice the virial radius, referred to as Sgr A* run. The latter corresponds to the case with no missing baryons and is our fiducial case. The adopted parameters for the initial conditions are tabulated in \tref{tab:ic} and additional numerical details are summarized in \aref{a:ic}.

\section{Results}

We present the results of our matrix of runs starting from the set of initial conditions outlined above for the case of an M87* analog with a central SMBH with a mass of $M_{\rm BH} \sim 6.5 \times 10^9 \Msun$ and for the case of a Sgr A* analog central SMBH with a mass of $M_{\rm BH} \sim 4.3 \times 10^6 \Msun$. For each of these BH masses, we perform Runs A, B, C1, C2 and C3 varying accretion and feedback prescriptions as noted above. This provides us with a library of models to study and compare with EHT data. 

\begin{figure*}
    \centering
    \includegraphics[width=17cm]{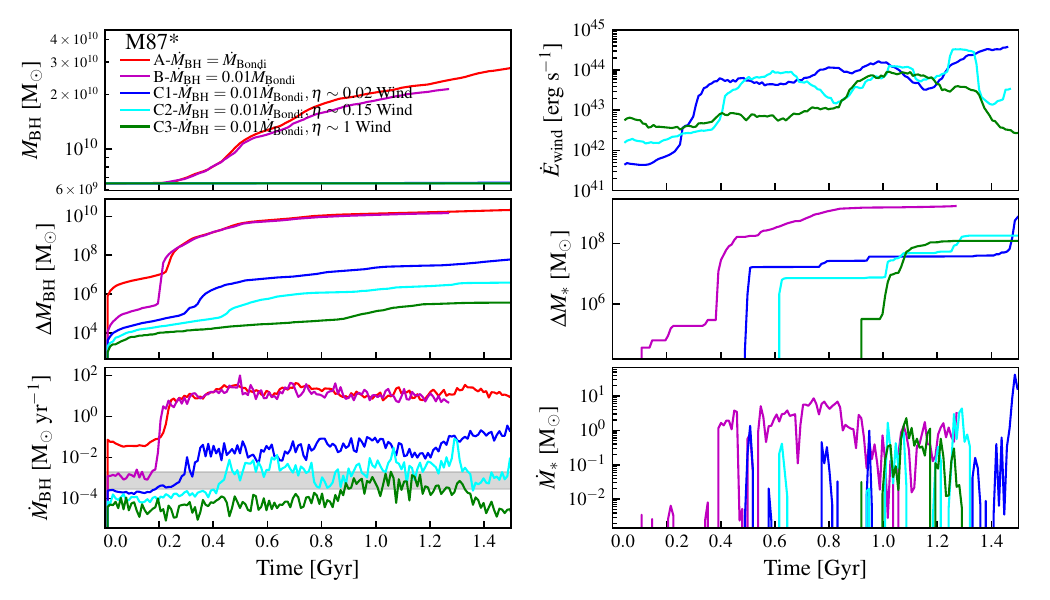}
    \caption{Results for the M87* analog runs: the BH mass; AGN wind energy flux (averaged over 100Myr); net increase in BH mass; net mass in stars formed; along with the corresponding BH accretion rates and star formation rates for Runs A (red curves), B (magenta curves), C1 (purple curves), C2 (blue curves)\& C3 (green curves) . The grey shaded region marks the BH accretion rate estimated from EHT data \citep{EHTm87_8_2021}. Without AGN feedback, both Runs A and B, exhibit runaway BH growth, even when the accretion rate is suppressed to 0.01 of the Bondi rate. As the AGN feedback efficiency is increased in Runs C1, C2 and C3, the corresponding BH accretion rates and star formation rates are lower. As seen a feedback efficiency ranging between $0.15-1$ (Runs C2 and C3) is required to match the EHT-inferred accretion rate. Note that Run A (red curve) has no star formation throughout the simulation, so it does not appear in the stellar mass and star formation rate panels.}
    \label{fig:M87}
\end{figure*}

\begin{figure*}
    \centering
    \includegraphics[width=17cm]{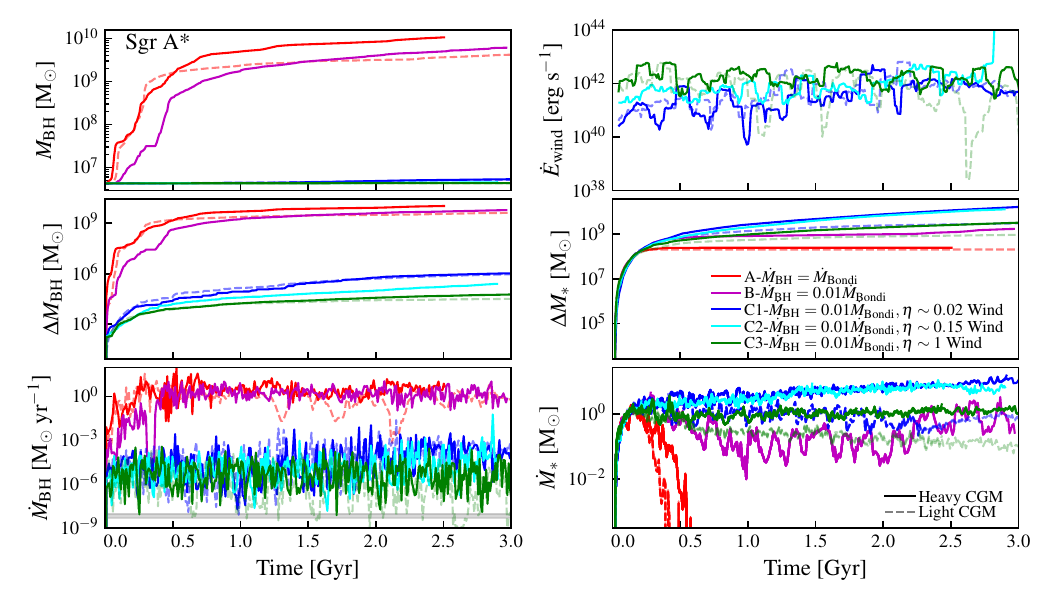}
    \caption{Results for the Sgr A* analog runs: the BH mass; AGN wind energy flux (averaged over 100Myr); net increase in BH mass; net stellar mass; along with the corresponding BH accretion rates and star formation rates for Runs A (red curves), B (magenta curves), C1 (purple curves), C2 (blue curves)\& C3 (green curves). The grey shaded region marks the BH accretion rate estimated EHT data \citep{EHTMW_5_2022}. Without AGN feedback both Runs A and B exhibit runaway BH growth, leading to rapid growth by orders of magnitude that occurs due to the accretion of gas that would have otherwise formed stars, thereby significantly lowering the star formation rate. The higher the feedback efficiency, seen in Runs C1, C2 and C3, the lower the corresponding BH accretion rate and star formation rate. A feedback efficiency of $\eta \sim 1$ (Run C3) or $\eta \sim 0.02$ (Run C1), assuming missing baryons) is required to suppress the star formation rate to $\lesssim 2 \, {\rm M_\odot}\,{\rm yr}^{-1}$ \citep[e.g.,][]{Chomiuk+MWSFR2011, Licquia+MWSFR2015, Elia+MWSFR2022}. A feedback efficiency of $\sim 1$, combined with a low CGM mass (grey dashed line) and missing baryons, is likely necessary to maintain a BH accretion rate that is consistent with the EHT data for an extended period.}
    \label{fig:mw}
\end{figure*}

\subsection{BH mass and accretion rates}

Starting with the current BH masses corresponding to the values of M87* and Sgr A*, in \fref{fig:M87} and \fref{fig:mw} we show the evolution of BH mass, stellar mass, AGN wind energy flux, as well as the corresponding BH accretion rates and star formation rates for our runs. The baryonic mass includes the gas mass, stellar mass, and BH mass, and the cooling flow would add to this baryonic reservoir. In the two runs without AGN feedback, Runs A and B, the BH grows extremely rapidly in all cases exceeding the present day mass of both the Sgr A* and M87* BHs. Suppressing the accretion rate to 1\% of the Bondi accretion rate without including any feedback in Run B, does not significantly mitigate this overgrowth. Comparing Runs A and B, we note that the net effect of no AGN feedback is to merely delay the overgrowth of BHs by a few Myr. The reason is that in the suppressed Bondi accretion rate case, namely Run B, 99\% of gas lingers and accumulates in the vicinity of the BH as only 1\% is accreted initially. This accumulation however, enhances the density and eventually boosts the mass accretion rate, leading to dramatic growth with a small delay. In the absence of AGN feedback, in Runs A and B for both BH masses, we swiftly exceed the present day mass within 0.1-0.2 Gyr. Meanwhile, for both BH masses in Runs C1, C2 and C3, there is no mass growth for upto 1.5 Gyr.

Including AGN feedback (in Runs C1, C2 and C3) has a much larger impact on BH accretion over longer periods of time. For both the M87* and Sgr A* analog cases, as long as there is some amount of AGN feedback with an efficiency $\eta > 0.02$, the accretion rate and BH mass do not deviate much from their initial values. Different feedback efficiencies however, do affect BH accretion rates. The higher the feedback efficiency, the lower the mass accreted by the BH. For both the M87* and Sgr A* analog cases, in the corresponding Runs C1, C2 \& C3, the net injected energy, which is proportional to the net accreted mass times the feedback efficiency ($ \eta \Delta M_{\rm BH}$), remains roughly constant, consistent with the feedback regulation scenario proposed for high redshift atomic cooling halos in \cite{Su+2023}.\footnote{Self-regulation operates for atomic cooling halos for the case of isotropic AGN-inflated bubbles.}

For both the M87* and Sgr A* analog cases, we overplot the range of BH accretion rates estimated from EHT data: $3-20 \times 10^{-4} {\rm M_\odot} \, {\rm yr}^{-1}$ for M87* \citep{EHTm87_8_2021} and $5.2-9.5 \times 10^{-9}\, {\rm M_\odot} \, {\rm yr}^{-1}$ for the Sgr A* \citep{EHTMW_5_2022} as the gray band. For the M87* analog case, a feedback efficiency of $\eta = 0.15-1$ is required to maintain the accretion rate at the observed value for an extended period of time (more than a few hundred Myr). It is clear that the feedback efficiency, $\eta \sim 0.02$, obtained from our previous bridging-scale work with magnetic fields and a non-spinning BH \citep{Cho+2023,Cho+2024}, is too low to sustain and reproduce the BH accretion rate inferred from the EHT data. Some degree of BH spin appears to be necessary to match the EHT data which also places limits on BH spin (the best-fit GRMHD model for M87* yields a non-zero spin of $|a| > 0.5$ \citep{EHTm87_8_2021} and $a = 0.94$ for Sgr A* \citep{EHTMW8+2024}). Despite the short-term variability in BH accretion rates seen for both BH masses in the simulations, we infer a typical feedback duty cycle of a few hundred ${\rm Myr}$.

For the Sgr A* analog case, even a feedback efficiency of $\eta \sim 1$ (Run C3), is insufficient to match the observed extremely low accretion rate of $< 10^{-8} {\rm M_\odot}\, {\rm yr}^{-1}$, assuming no missing baryons in the CGM. However, under the assumption of a more observationally motivated CGM profile (the Light CGM case), with a feedback efficiency of $\eta \sim 1$ in Run C3 can, we see that from time to time, a sharp drop in the accretion rate down to values of $< 10^{-8} {\rm M_\odot}\, {\rm yr}^{-1}$ occur matching the observationally inferred value. Upon extending the run times of the higher efficiency Light CGM case, we find that such sharp drops in the accretion rate become increasingly frequent and persist for longer periods after the initial 1.0 Gyr or so. 

Intriguingly, constraints from X-ray wavelengths, from Fermi and eROSITA bubbles \citep{Yang+2022Nature} provide significantly higher estimates for the accretion rate for Sgr A* reaching $\sim 2 {\rm M_\odot}\,{\rm yr}^{-1}$ and our simulation results for Runs A and B are consistent with that value and the rest of the runs lie several orders of magnitude lower. 

\subsection{The star formation rate}

As expected, AGN feedback also significantly affects star formation rates as seen in \fref{fig:M87} \& \fref{fig:mw}. Generally speaking, the higher the feedback efficiencies, the lower the star formation rates for both the M87* and Sgr A* analog cases as shown with the Runs C1-C3. For the M87* analog case, all runs with AGN feedback exhibit star formation rates of $\ll 1 {\rm M_\odot} \, {\rm yr}^{-1}$ for the majority of the simulation time, which is consistent with the very low observed value \citep[e.g.,][]{Tan+2005, CookConroy+2020}. Without AGN feedback and with accretion at the Bondi rate (Run A), the BH swallows all of the inflowing gas in its vicinity before any stars can form, resulting in very scarce star formation. The gas in this case, preferentially powers BH accretion rather than star formation. On the other hand, suppressed Bondi accretion without AGN feedback (Run B) allows some of accumulated gas sufficient time to form stars before being accreted by the BH. It appears in this case, that a star formation rate ranging between  $1-10\, {\rm M_\odot} \, {\rm yr}^{-1}$ can be sustained for several Myr, which is somewhat higher than what current observations suggest.

For the Sgr A* analog case, assuming no missing baryons, only the highest feedback efficiency case with $\eta \sim 1$ (Run C3) suppresses the star formation rate to $\lesssim 2\, {\rm M_\odot}\, {\rm yr}^{-1}$, in agreement with observations \citep[e.g.,][]{Chomiuk+MWSFR2011, Licquia+MWSFR2015, Elia+MWSFR2022}. A much lower AGN feedback efficiency of $\eta \sim 0.02$ (as in Run C1) is sufficient to suppress star formation to $\lesssim 2 \,{\rm M_\odot} \, {\rm yr}^{-1}$ for the lighter CGM case (dashed purple line)). Run C3 with the lighter CGM case, and a higher feedback efficiency (dashed green line), meanwhile, results in a star formation rate of $\lesssim 0.6 {\rm M_{\odot}} \, {\rm yr}^{-1}$, which may be slightly over quenched compared to the observed MW value. Interestingly, in the Sgr A* analog cases studied without AGN feedback (Runs A and B), the star formation rate is significantly lower than in the runs including AGN feedback. This occurs because, in these runs, regardless of full or suppressed Bondi accretion, the BH mass grows to $>10^9 {\rm M_{\odot}}$, more than 3 orders of magnitude larger than its initial value consuming all the gas. The significantly larger BH consumes much of the gas leading to extremely high BH accretion rates removing gas rapidly that would otherwise form stars, resulting in a reduced star formation rate. In particular, in the case of full Bondi accretion, star formation is quenched after $\sim 0.5$ Gyr. We note that the BH accretion rate and star formation rates for the Sgr A* analog case depend largely on the detailed initial conditions adopted for the properties of ISM and CGM. To precisely match the BH accretion rate and the star formation rate, over an extended period of time would require more detailed modeling and fine-tuning of the initial conditions of the gas distribution. However, based on our set of runs, we can still robustly conclude that feedback efficiencies much higher than the percent level -- requiring BH spin -- appear to be required to match the data. We leave the more detailed modeling of gas initial conditions for future work.

\subsection{Gas profiles}\label{sec:gas_profile}

\begin{figure*}
    \centering
    \includegraphics[width=17cm]{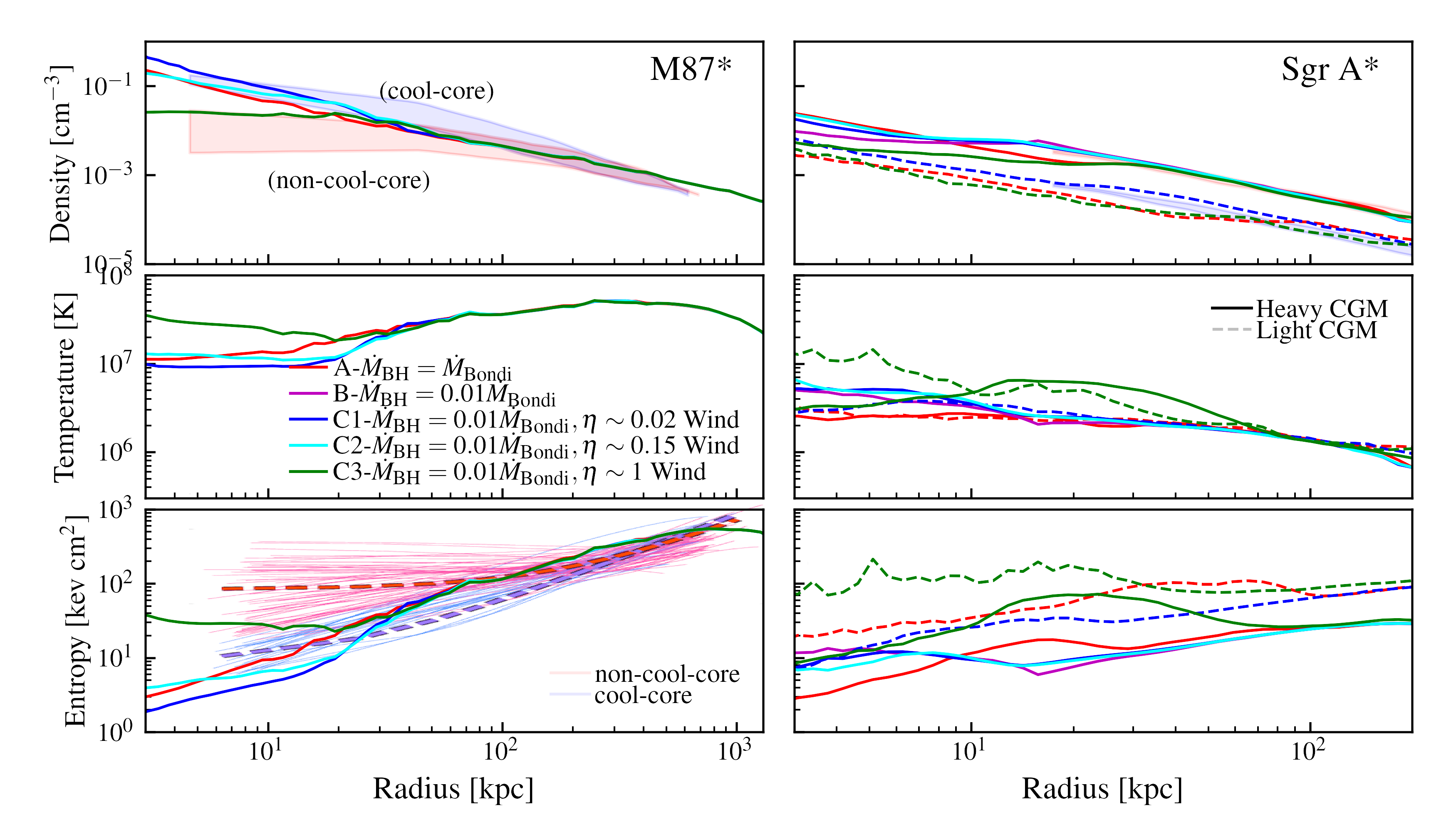}
    \caption{Density, temperature, and entropy profiles for the M87* and Sgr A* analog runs, averaged from 1.45 to 1.5 Gyr. The M87* analog profiles are X-ray luminosity-weighted (0.5-7 keV), while the Sgr A* analog profiles are mass-weighted for gas with $T > 10^5$ K. Shaded regions in the M87* analog panels represent observed cool-core (blue) and non-cool-core (red) cluster profiles \citep{McDonald+cluster2013} (scaled to account for the halo mass differences), and in the Sgr A* analog panels, the Milky Way profiles \citep{Miller+MW2015} (blue and red, the latter assuming no missing baryons). The M87* analog run with $\eta \sim 1$ (Run C3) yields profiles that resemble non-cool-core clusters, while the other runs resemble cool-core clusters.}
   \label{fig:etd}
\end{figure*}

\fref{fig:etd} presents the density, temperature, and entropy profiles for both the M87* and Sgr A* analog runs, averaged over the period from 1.45 to 1.5 Gyr. Since the virial temperature of the gas in the MW is $\gtrsim 10^6$ K and it is not X-ray bright, 
we employ mass-weighted profiles for gas hotter than $10^5$ K, rather than X-ray-luminosity-weighted profiles for Sgr A* analog runs. For the M87* analog runs, we use X-ray luminosity-weighted values in the 0.5-7 keV range. 
In the Sgr A* analog density panel, the narrow shaded regions in blue and red represent the observed density profile of the MW, under the assumption of missing baryons in the CGM (blue, as noted by \cite{Miller+MW2015}), and no missing baryons (red). For the M87* analog,  density and entropy panels, the over-plotted shaded regions indicate observational profiles derived for cool-core (blue) bearing low-entropy core and non-cool-core (red) clusters having a flat entropy profile extending to the core \citep{McDonald+cluster2013}, scaled appropriately for M87.

Interestingly, for the M87* analog runs explored here, the resulting gas properties all fall within the observed range marked by the shaded regions. AGN feedback increases the gas temperature and decreases the gas density. The highest feedback efficiency of $\eta \sim 1$ in Run C3 results in flattened gas entropy and density profiles that resemble observed non-cool-core clusters (shown as a red shaded region), while all the other runs more closely resemble cool-core clusters.

In the Sgr A* analog runs, a similar trend is observed, wherein AGN feedback raises the gas temperature and lowers the gas density. However, none of the AGN feedback models explored in this work are able to produce winds strong enough to reduce the gas density profile from the one assuming no missing baryons (red) to match the observed profile in \cite{Miller+MW2015} (blue). Even the most aggressive AGN feedback model assumed here is unable to dilute and redistribute baryons to account for the missing baryons. Achieving this likely requires energy fluxes on the order of $\sim 5 \times 10^{42} \, {\rm erg \, s}^{-1}$ \citep{Su+JetHaloMass2024}. In all our Sgr A* analog runs, the average energy flux is below $\sim  10^{42} \, {\rm erg \, s}^{-1}$, even for the highest feedback efficiency of $\eta \sim 1$ studied here. This suggests that stronger prior episodes of AGN feedback, possibly driven by mergers, are likely required to reconcile the missing baryon budget in the MW as noted in \cite{Miller+MW2015}.

Additionally, we observe a noticeable suppression of gas density within 30 kpc in the Sgr A* analog Run A with full Bondi accretion and no AGN feedback. This is consistent with the BH accreting excessively and the amplified effect of stellar feedback in low-density regions.

\subsection{Morphology of the systems}
\begin{figure*}
    \centering
    \includegraphics[width=17cm]{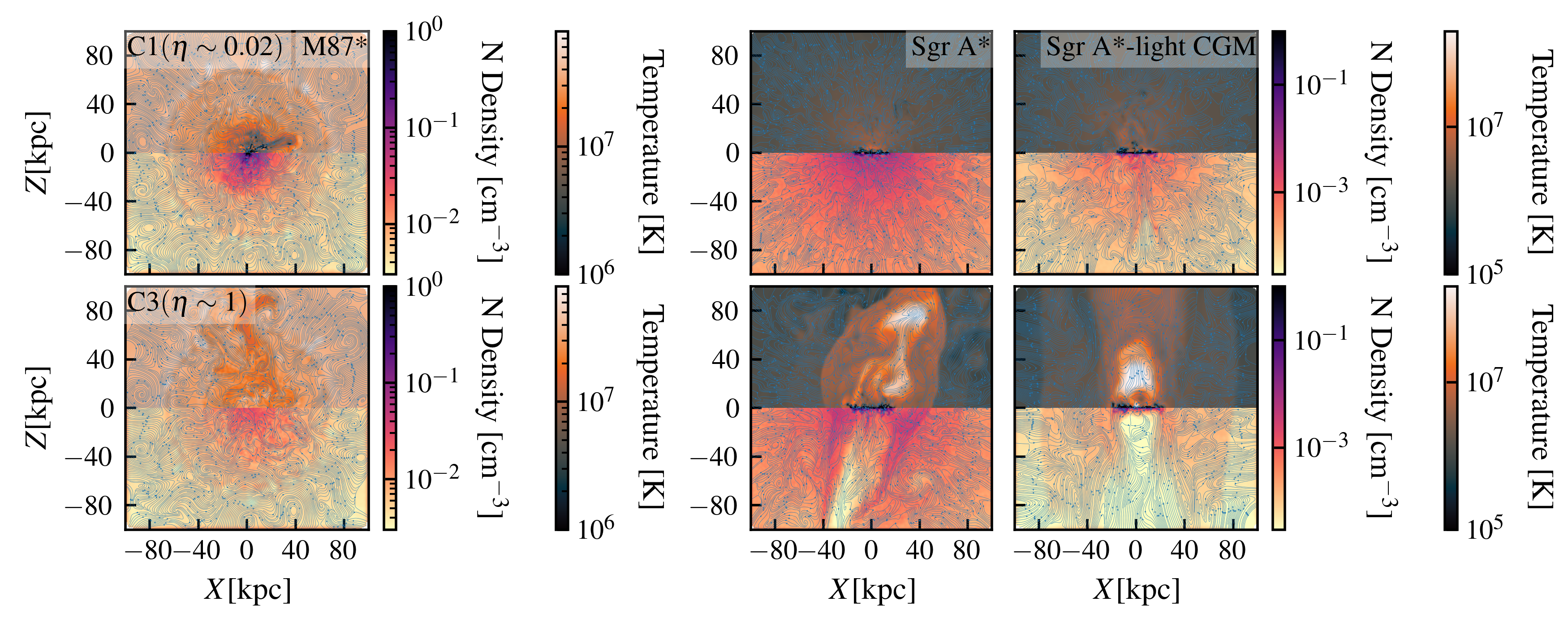}
    \caption{The gas temperature (upper half) and number density (lower half) morphology in a central $|y|<$ 2 kpc slice for the runs with feedback efficiencies of $\eta \sim 0.02$ (C1, upper row) and $\eta \sim 1$ (C3, lower row) at 1.5 Gyr. The blue lines represent the magnetic field lines. In the Sgr A* analog runs C3, with $\eta \sim 1$, the AGN cocoon propagates out to 80 kpc.}
    \label{fig:morph}
\end{figure*}
\fref{fig:morph} shows the gas morphology in a central $|y|<$ 2 kpc slice for the runs with feedback efficiencies of $\eta \sim 0.02$ (upper row) and $\eta \sim 1$ (lower row), respectively. We plot the temperature in the upper half of the plot,
and the number density in the lower half. Consistent with \sref{sec:gas_profile}, higher feedback efficiency results in lower gas density and higher gas temperature.
In the Sgr A* analog cases Run C3 with $\eta \sim 1$, a strong AGN-inflated cocoon is clearly visible, extending up to $>80$ kpc. Despite the initially isotropic injection, the cocoon is largely collimated by the ISM gas disk.

\section{Conclusions \& Discussion}
In this work, we test the effects of the suppression of Bondi accretion and AGN feedback for the central supermassive BHs hosted in M87 and MW-like halos using high-resolution, non-cosmological magnetohydrodynamic simulations with the FIRE-2 stellar feedback model. Our key findings are: 
\begin{itemize} 
\item{With no AGN feedback and suppressed accretion compared to the Bondi rate - $0.01\dot{M}_{B}$ (Run B) - the cooling flow is not stopped and BH overgrowth appears to be inevitable. As gas continues to flow into the vicinity of the BH, the unaccreted gas increases in density, eventually raising the Bondi accretion rate. The initial suppression merely delays BH overgrowth.} 
\item{When we include AGN feedback with efficiency $>0.02$ combined with suppressed Bondi accretion, the BH accretion rate is significantly reduced. Over several Gyr, the BH mass changes only by a few percent.} 
\item{Higher feedback efficiencies result in stronger suppression of cooling flows and lower BH accretion rates. For both M87* and the Sgr A* analogs, self-regulation maintains the average energy flux from the BH at a consistent level.}
\item {For the M87* analog, an AGN feedback efficiency of 0.15--1 (Runs C2, C3) is required to sustain a BH accretion rate of $10^{-4}$--$10^{-3} ~{\rm M_\odot} {\rm yr}^{-1}$, consistent with EHT data \citep{EHTm87_8_2021}.} 
\item{For the Sgr A* analog, a feedback efficiency of $\sim1$ is required to reduce star formation to $\lesssim 2 {\rm M_\odot} {\rm yr}^{-1}$, assuming no missing baryons. With a lighter CGM profile, as suggested by \cite{Miller+MW2013, Miller+MW2015}, a feedback efficiency of $\gtrsim 0.15$ is sufficient to reach a similar level of suppression of star formation.}
\item{Even with the highest feedback efficiency of $\sim1$ (in Run C3) and assuming a light CGM, the BH accretion rate for the Sgr A* analog remains around $10^{-6} {\rm M_\odot} {\rm yr}^{-1}$, nearly two orders of magnitude higher than inferred from EHT data \citep{EHTMW_5_2022}. However, after 2 Gyr, we see brief periods during which the BH accretion rate drops to $\lesssim 10^{-8} {\rm M_\odot} {\rm yr}^{-1}$, periodically, leading to matching the EHT result. This suggests that the current low accretion rate of Sgr A*, likely reflects a quiescent phase in its rapidly varying evolution. }
\item{In both the M87* and Sgr A* analog simulations, an AGN feedback efficiency of $>0.15$ is always required to achieve reasonable BH accretion rates and star formation suppression as observed. This level of feedback efficiency requires non-zero BH spin. For the Sgr A* analog, additionally, a lighter CGM profile, consistent with missing baryons, appears to also be necessary.}
\end{itemize}
We detail several caveats of our current analysis and possible improvements in \aref{a:numerical} that are beyond the scope of this work.

\section*{Acknowledgements}
 
KS, PN, HC, and RN were partially supported by the Black Hole Initiative at Harvard University, which is funded by the Gordon and Betty Moore Foundation grant 8273, and the John Templeton Foundation grant 61497. The opinions expressed in this publication are those of the authors and do not necessarily reflect the views of the Moore or Templeton Foundation. 
This work was performed in part at Aspen Center for Physics, which is supported by National Science Foundation grant PHY-2210452.
The simulations are done on Frontera with allocation AST22010, and Bridges-2 with Access allocations TG-PHY220027 \& TGPHY220047. 
Support for PFH was provided by NSF Research Grants 1911233, 20009234, 2108318, NSF CA-REER grant 1455342, NASA grants 80NSSC18K0562, HST-AR-15800.   
This research also used resources provided to BP by the Los Alamos National Laboratory Institutional Computing Program, which is supported by the U.S. Department of Energy National Nuclear Security Administration under Contract No. 89233218CNA000001.  We thank FIRE collaboration for useful discussions.

\appendix
\section{Numerical details}
\subsection{Stellar Physics}\label{a:sfb}
 All our simulations have the FIRE-2 implementation physical treatments of the ISM, star formation, and stellar feedback, the details of which are given in \citet{Hopkins+FIRE2018} along with extensive numerical tests.  Cooling is followed from $10-10^{10}$K, including the effects of photo-electric and photo-ionization heating, collisional, Compton, fine structure, recombination, atomic, and molecular cooling. 

Star formation is treated via a sink particle method, allowed only in molecular, self-shielding, locally self-gravitating gas above a density $n>100\,{\rm cm^{-3}}$ \citep{Hopkins+SFC2013}. Star particles, once formed, are treated as a single stellar population with metallicity inherited from their parent gas particle at formation. All feedback rates (SNe and mass-loss rates, spectra, etc.) and strengths are IMF-averaged values calculated from {\small STARBURST99} \citep{Leitherer+Starburst99_1999} with a \citet{Kroupa+IMF2002} IMF. The stellar feedback model includes (1) Radiative feedback, including photo-ionization and photo-electric heating, as well as single and multiple-scattering radiation pressure tracked in five bands  (ionizing, FUV, NUV, optical-NIR, IR), (2) OB and AGB winds, resulting in continuous stellar mass loss and injection of mass,  metals, energy, and momentum  (3) Type II and Type Ia SNe (including both prompt and delayed populations) occurring according to tabulated rates and injecting the appropriate mass, metals, momentum, and energy to the surrounding gas. All the simulations run also include MHD.

\subsection{AGN Wind Implementation}\label{a:agn}

We launch the wind with a particle spawning method \citep{Torrey+2020,Su+2021,Weinberger+2023,Su+JetHaloMass2024}, which creates new gas cells (resolution elements) from the central BH. With this method, we have better control of the jet properties as the launching is less dependent on the neighbor-finding results. We can also enforce a higher resolution for the jet elements.  The form of feedback in this paper is similar to \cite{Torrey+2020}, which studied the effects of broad absorption line (BAL) wind feedback on disk galaxies. 
The spawned gas particles have a mass resolution of labeled in \tref{tab:ic} and are forbidden to de-refine (merge into a regular gas element) before they decelerate to 10\% of the launch velocity. Two particles are spawned in opposite directions at the same time when the accumulated jet mass flux reaches twice the target spawned particle mass, so linear momentum is always exactly conserved. Initially, the spawned particle is randomly placed on a sphere with a radius of $r_0$, which is either 10 pc or half the distance between the BH and the closest gas particle, whichever is smaller.  The initial velocity is at the radial direction.  The initial magnetic field of the spawned wind particle is $<10^{-4} \, G$ in the z-direction, and the initial temperature of the AGN wind is $10^7 , K$. Both the thermal and magnetic energy fluxes are subdominant to the kinetic energy flux.

\subsection{Initial conditions}\label{a:ic}

Initial conditions for the galaxy-scale simulations resemble the resemble an M87*-like host galaxy \citep{2008MNRAS.388.1062C, 2012MNRAS.421..635F, 2016MNRAS.457..421O} and the Sgr A* host Milky Way \citep{Miller+MW2013, Miller+MW2015, Gupta+2017}. We include two initial conditions for Sgr A* host Milky Way each assuming different CGM masses: one follows the CGM profile from \cite{Miller+MW2015} (Sgr A*-light CGM), while the other uses a similar profile but assumes a universal baryonic fraction (0.16) within twice the virial radius. The latter is the fiducial case to test whether any feedback models can account for the missing baryons suggested in \cite{Miller+MW2015}.

The initial conditions for the dark matter (DM) halo, stellar bulge, BH, and gas halo are initialized following the procedures in \cite{1999MNRAS.307..162S}, \cite{2000MNRAS.312..859S}, and \cite{Su+2019, Su+2020, Su+2021, Su+JetHaloMass2024}. We begin with a spherical, isotropic DM halo with an NFW profile \citep{1996ApJ...462..563N}, having a halo mass (within $R_{200}$) of $(1.5 \times 10^{12}, 1.97 \times 10^{14})\,{\rm M_\odot}$ and a scale radius of (20, 448) kpc for the Sgr A* and M87* analog runs, respectively. The stellar bulge follows a \cite{1990ApJ...356..359H} profile with masses of $(1.5 \times 10^{11}, 6.87 \times 10^{11})\, {\rm M_\odot}$ and scale radii of (1, 3.97) kpc for the Sgr A* and M87* analog runs. The BH has a mass of $M_{\rm BH} = (4.3 \times 10^6, 6.5 \times 10^9)\, {\rm M_\odot}$ for Sgr A*  and  M87*. The gas halo is in hydrostatic equilibrium and follows a $\beta$-profile with a mass (within $R_{200}$) of $(1.6 \times 10^{11}, 4.6 \times 10^{10}, 4.7 \times 10^{13})\,{\rm M_\odot}$, $\beta = (0.5, 0.5, 0.33)$, and scale radii of (20, 20, 0.93) kpc for the Sgr A* , Sgr A* -light CGM, and M87* runs, respectively. The hydrostatic profile evolves into a rotating cooling flow solution \citep{Stern+coolingflow2024} as the simulations progress.

The gas in the M87* analogue halo rotates at $0.1\Omega_K$, where $\Omega_K$ is the Keplerian angular velocity, and it is primarily supported by thermal pressure, as expected in such massive halos. The Sgr A* host Milky Way also start with a rotation of $0.3\Omega_K$, but with a declining rotational velocity of $v_\phi \propto r^{-1}$ for $r > 30$ kpc, creating a profile with roughly constant angular momentum. For the Sgr A* analogue runs, we also assume exponential, rotation-supported gas and stellar disks with scale lengths of 6 kpc and 3 kpc, respectively, and a scale height of 0.3 kpc for both.

The metallicity profiles follow $Z_\odot (Z_{\rm out} + (1 - Z_{\rm out}) / (1 + (r / R_c)^{\beta_z}))$, where $Z_{\rm out} = (0.05, 0.26)$, $R_c = (20, 64)$ kpc, and $\beta_z = (1.5, 2)$ for the Sgr A* and M87* analog runs. For the M87* analog runs, we assume a temperature floor ($T>3e6K$) for gas within 1 kpc to maximally match the gas thermal properties described in \citep{Cho+2024}. Since the Milky Way has a much lower virial temperature and a disk of ISM, no temperature floor is applied for the Sgr A* analogue runs.

In all runs, the magnetic fields decay from an initial poloidal configuration with $\beta \sim 1$ within 1 kpc, transitioning to toroidal fields with $\beta = 1000$ beyond 10 kpc. The poloidal magnetic field follows the expression in \cite{Cho+2024}, with the overall magnitude adjusted for the different initial conditions to maintain roughly the same maximum $\beta$ within 1 kpc.

The Sgr A* analog runs use a uniform gas resolution of 8,000 ${\rm M_\odot}$ within 244 kpc. The M87* analogue runs have a maximum gas mass resolution of $2 \times 10^4 {\rm M_\odot}$. A hierarchical super-Lagrangian refinement scheme \citep[e.g.,][and our previous series of papers]{AnglesAlcazar+Zoom_2021,Su+2023,Su+JetHaloMass2024} is used to achieve the maximum mass resolution in the core region, with resolution decreasing as a function of radius ($r_{\rm 3d}$), proportional to $r_{\rm 3d}$, down to $\sim 2 \times 10^6 {\rm M}{\odot}$. The highest resolution is achieved in regions where $r{\rm 3d}$ is less than 10 kpc.

\subsection{Possible future improvements and current limitations}\label{a:numerical}

\subsubsection{Scaling for BH accretion rate}

In this work, we employ a fixed accretion fraction, $\dot{M}_{\rm BH} \sim 0.01 \dot{M}_B$, throughout our simulations. However, \cite{Cho+2024} demonstrate that $\dot{M}_{\rm BH}/\dot{M}_B \approx \left( R_B / 6 \, r_g \right)^{-0.5}$ \cite[see also][]{Guo+2024}. In our galaxy simulations, we can, in principle, estimate the real-time $R_B / r_g$, according to the black hole mass and gas temperature, and adjust the accretion rate accordingly. Following this approach, instead of applying the same accretion rate relative to Bondi accretion for both the Sgr A* and M87*, the accretion rate can be scaled as needed. However, we do not expect this to significantly alter our results, particularly for our M87* analog runs.

For the M87* analog simulations, we implement a temperature floor within 1 kpc of the black hole. Given this, and the relatively long cooling time for gas at its virial temperature, $T > 10^7$ K, the temperature around the black hole remains largely stable throughout the simulation. The Bondi radius should, therefore, also be similar to the value in \cite{Cho+2024}.

For the Sgr A* analog case, we did not apply a temperature floor. However, relative velocity term $\Delta v$ in the Bondi accretion expression (\Eqref{eq:bondi}), which effectively measures the turbulent velocity in the vicinity of the black hole, remains  $\gtrsim30$ km s$^{-1}$. As a result, the corresponding effective Bondi radius, $R_B \propto GM / (c_s^2 + \Delta v^2)$, also does not too significantly during the simulation. However, due to the overall lower ambient temperature, implying a smaller $R_B / r_g$, the actual ratio $\dot{M}_{\rm BH}/\dot{M}_B$ may exceed 0.02 by a factor of about $\sim 3$. We observe that self-regulation maintains the average output energy flux at approximately the same level. Given that
\begin{align}
\eta=\frac{f_{\rm mass, wind}V^2_{\rm wind}}{2 f_{\rm mass, acc}c^2} &= \frac{(1-f_{\rm mass, acc})V^2_{\rm wind}}{2 f_{\rm mass, acc}c^2} \sim \frac{V^2_{\rm wind}}{2 f_{\rm mass, acc}c^2}
\end{align}
assuming $f_{\rm mass, acc} + f_{\rm mass, BH} = 1$ and $f_{\rm mass, acc} \ll 1$, we also have
\begin{align}\label{eq:reg}
\langle\dot{m}_{\rm BH}\rangle&\sim \frac{\langle\dot{E}_{\rm wind}\rangle}{\eta c^2}\sim \frac{2\langle\dot{E}_{\rm wind\rangle} f_{\rm mass, acc}}{ f_{\rm mass, wind}V_{\rm wind}^2}\sim \frac{2\langle\dot{E}_{\rm wind\rangle} f_{\rm mass, acc}}{ V_{\rm wind}^2}
\end{align}
This implies that the black hole accretion rate we obtain could be underestimated by a factor of approximately $\sim 3$, assuming the wind velocity remains unchanged.

\subsubsection{Resolution effects}

The current resolution is consistent with our previous work \citep{Su+2021, Su+JetHaloMass2024}, which tested a series of fixed-flux winds and jets. However, for the current study involving live accretion, the resolution effect can influence the estimation of accretion rates. In the M87* analog runs, the Bondi radius, $\sim 30$ pc, is marginally resolved. In the Sgr A* analog case, where the Bondi radius is $\sim 0.02$ pc, it is far from resolved. In the current implementation, we directly use the density, temperature, and relative velocity from the neighborhood finding result in the Bondi radius expression (\Eqref{eq:bondi}). Depending on how the density and temperature scale outside the Bondi radius, this may affect the overall estimate. However, due to the self-regulation described in \Eqref{eq:reg}, it is  $f_{\rm mass, acc}$ and  $f_{\rm mass, wind}$ the parameters $f_{\rm mass, acc}$ and $f_{\rm mass, wind}$ ultimately determine the black hole accretion rate. Any overall factor multiplying the Bondi accretion rate estimate will cancel out. Therefore, we do not expect the resolution effect to qualitatively change our results, as long as self-regulation holds.

To rigorously verify this, super-Lagrangian refinement down to the Bondi radius scale, following \cite{AnglesAlcazar+Zoom_2021, Hopkins+ForgeInFIRE1+2024}, is needed which is beyond the scope of our current work. 

\subsubsection{Geometrical effects}

Another key limitation of our current study is the adoption of isotropic AGN winds, where most of the energy is in kinetic form. This form of feedback follows results from \cite{Cho+2023, Cho+2024}, which arise purely due to magnetic fields without the influence of BH spin. These results have only been tested in an M87*-like setup using the novel multi-zone GRMHD simulations thus far. For consistency, we apply the same form of feedback in our Sgr A* analog runs. 

In reality, for higher feedback efficiencies, which may require BH spin, the resulting feedback could become more collimated, resembling observed AGN jets. With a more elongated cocoon, the energy required to counteract the gas inflow may be even higher \citep{Su+2023,Su+highz2024}. If the jet cocoon remains very narrow up to the cooling radius, it may completely fail to regulate the cooling flow and quench star formation and black hole growth \citep{Su+2021,Su+JetHaloMass2024}. The impact of such geometrical effects is also left for future exploration.

\bibliography{bridging_scales}{}
\bibliographystyle{aasjournal}


\end{CJK}
\end{document}